\documentclass[twocolumn,prl,showpacs,amsmath,amssymb]{revtex4}

\usepackage{graphicx}

\def\be{\begin{equation}}
\def\ee{\end{equation}}
\def\bea{\begin{eqnarray}}
\def\eea{\end{eqnarray}}
\def\bma{\begin{mathletters}}
\def\ema{\end{mathletters}}

\def\0{\overline{0}}

\def\q0{\underline{0}}

\def\L{{\cal L}}

\def\tr{\mbox{tr}}
\def\one{\leavevmode\hbox{\small1\normalsize\kern-.33em1}}
\def\compl{\begin{picture}(8,8)\put(0,0){C}\put(3,0.3){\line(0,1){7}}\end{picture}}

\def\bra#1{\langle#1|} \def\ket#1{|#1\rangle}
\def\braket#1#2{\langle#1|#2\rangle}

\def\proj#1{\ket{#1}\!\bra{#1}}

\begin{document}

\title{
Asymptotic quantum cloning is state estimation }

\author{Joonwoo Bae and Antonio Ac\'\i n}

\affiliation{ICFO-Institut de Ci\`encies Fot\`oniques,
Mediterranean Technology Park, 08860 Castelldefels (Barcelona), Spain}

\date{\today}


\begin{abstract}

The impossibility of perfect cloning and state estimation are two
fundamental results in Quantum Mechanics. It has been conjectured
that quantum cloning becomes equivalent to state estimation in the
asymptotic regime where the number of clones tends to infinity. We
prove this conjecture using two known results of Quantum
Information Theory: the monogamy of quantum correlations and the
properties of entanglement breaking channels.

\end{abstract}


\maketitle


The impossibility of perfect state estimation is a major
consequence of the nonorthogonality of quantum states: \emph{the
state of a single quantum system cannot be perfectly  measured}.
In other words, a measurement on a system in order to acquire
information on its quantum state perturbs the system itself. The
full reconstruction of the state is only possible by computing
statistical averages of different observables on a large number of
identically prepared systems. Thus, any measurement at the
single-copy level only provides partial information.

The fact that state estimation is in general imperfect leads in a
natural way to the problem of building \emph{optimal
measurements}. Being a perfect reconstruction impossible, it is
relevant to find the measurement strategy that maximizes the gain
of information about the unknown state. A standard approach to
this problem in Quantum Information Theory (QIT) is to quantify
the quality of a measurement by means of the so-called
\emph{fidelity} \cite{MP}. This quantity is defined as follows.
Consider the situation in which a quantum state $\ket{\psi}$ is
chosen from the ensemble $\{p_i,\ket{\psi_i}\}$, i.e. $\ket{\psi}$
can be equal to $\ket{\psi_i}$ with probability $p_i$. A
measurement, defined by $N_M$ positive operators, $M_j\geq 0$,
summing up to the identity, $\sum_j M_j=\one$, is applied on this
unknown state. For each obtained outcome $j$, a guess
$\ket{\phi_j}$ for the input state is made. The overlap between
the guessed state and the input state,
$|\braket{\psi_i}{\phi_j}|^2$, quantifies the quality of the
estimation process. The averaged fidelity of the measurement then
reads
\begin{equation}\label{avfidmeas}
    \bar F_M=\sum_{i,j}
    p_i\, \tr(M_j\proj{\psi_i})\,|\braket{\psi_i}{\phi_j}|^2 .
\end{equation}
A measurement is optimal according to the fidelity criterion when
it provides the largest possible value of $\bar F_M$, denoted in
what follows by $F_M$.

The No-cloning theorem \cite{WZ}, one of the cornerstones of QIT
\cite{reviews}, represents another known consequence of the
nonorthogonality of quantum states. It proves that \emph{given a
quantum system in an unknown state $\ket{\psi}$, it is impossible
to design a device producing two identical copies,
$\ket{\psi}\ket{\psi}$}. Indeed, two nonorthogonal quantum states
suffice to prove the no-cloning theorem.

As it happens for state estimation, the impossibility of perfect
cloning leads to the characterization of \emph{optimal cloning
machines} \cite{BH}. In this case, one looks for the quantum map
$\L$ that, given a state $\ket{\psi}$ chosen from an ensemble
$\{p_i,\ket{\psi_i}\}$ in $\compl^d$, produces a state
$\L(\psi)=\rho_{C_1\ldots C_N}$ in $(\compl^d)^{\otimes N}$, such
that each individual clone $\rho_{C_k}=\tr_{\bar
k}(\rho_{C_1\ldots C_N})$ resembles as much as possible the input
state, where $\tr_{\bar k}$ denotes the trace with respect to all
the systems $C_1,\ldots,C_N$ but $C_k$. The average fidelity of
the cloning process is then
\begin{equation}\label{avfidcl}
    \bar F_C(N)=\sum_{i,k}
    p_i\, \frac{1}{N}\bra{\psi_i}\tr_{\bar k}\L(\psi_i)\ket{\psi_i} .
\end{equation}
The goal of the optimal machine is to maximize this quantity, this
optimal value being denoted by $F_C(N)$.

One can easily realize that the no-cloning theorem and the
impossibility of perfect state estimation are closely related. On
the one hand, if perfect state estimation was possible, one could
use it to prepare any number of clones of a given state, just by
measurement and preparation. On the other hand, if perfect cloning
was possible, one could perfectly estimate the unknown state of a
quantum system by preparing infinite clones of it and then
measuring them. Beyond these qualitative arguments, the connection
between state estimation and cloning was strengthened in
\cite{GM,BEM}. The results of these works suggested that
asymptotic cloning, i.e. the optimal cloning process when
$N\to\infty$, is equivalent to state estimation, in the sense that
\begin{equation}\label{conj}
    F_C=F_C(N\to\infty)=F_M .
\end{equation}
Later, this equality was rigorously shown to hold in the cases of
(i) universal cloning \cite{KW}, where the initial ensemble
consists of an arbitrary pure state in $\compl^d$, chosen with
uniform probability, and (ii) phase covariant qubit cloning
\cite{BCDM}, where the initial ensemble corresponds to a state in
$\compl^2$ lying on one of the equators of the Bloch sphere. Since
then, the validity of this equality for any ensemble has been
conjectured and, indeed, has been identified as one of the open
problems in QIT \cite{web}.

In this work, we show that the fidelities of optimal asymptotic
cloning and of state estimation are indeed equal for any initial
ensemble of pure states. Actually, we prove that \emph{asymptotic
cloning does effectively correspond to state estimation}, from
which the equality of the two fidelities trivially follows. The
proof of this equivalence is based on two known results of QIT:
the monogamy of quantum correlations and the properties of the
so-called entanglement breaking channels (EBC).


It is easy to prove that $F_M\leq F_C$. Indeed, given the initial
state $\ket{\psi}$, a possible asymptotic cloning map, not
necessarily optimal, consists of first applying state estimation
and then preparing infinite copies of the guessed state. It is
sometimes said that the opposite has to be true since ``asymptotic
cloning cannot represent a way of circumventing optimal state
estimation". As already mentioned in \cite{web}, this reasoning is
too naive, since it neglects the role correlations play in state
estimation. For instance, take the simplest case of universal
cloning of a qubit, i.e. a state in $\compl^2$ isotropically
distributed over the Bloch sphere. The optimal cloning machines
produces $N$ approximate clones pointing in the same direction in
the Bloch sphere as the input state, but with a shrunk Bloch
vector \cite{KW}. If the output of the asymptotic cloning machine
was in a product form, it would be possible to perfectly estimate
the direction of the local Bloch vector, whatever the shrinking
was. Then, a perfect estimation of the initial state would be
possible. And of course, after the perfect estimation one could
prepare an infinite number of perfect clones! This simple
reasoning shows that the correlations between the clones play an
important role in the discussion. Actually, it has recently been
shown that the correlations present in the output of the universal
cloning machine are the worst for the estimation of the reduced
density matrix \cite{rafael}.

As announced, the proof of the conjecture is based on two known
results of QIT: the monogamy of entanglement and the properties of
EBC. For the sake of completeness, we state here these results,
without proof.

\begin{figure}
  \includegraphics[width=8cm]{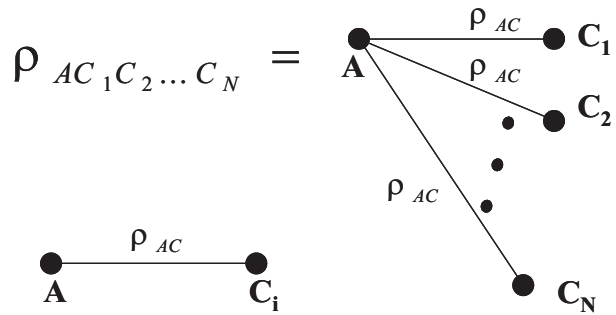}\\
  \caption{The state $\rho_{AC}$ is said to be $N$-shareable when there
  exists a global state $\rho_{AC_1\ldots C_N}$ such that
  the local state shared between $A$ and $C_i$
  is equal to $\rho_{AC}$, for all $i$.}\label{entshar}
\end{figure}

Quantum correlations, or entanglement, represent a monogamous
resource, in the sense that they cannot be arbitrarily shared. One
of the strongest results in this direction was obtained by Werner
in 1989 \cite{Werner}. There, it was shown that the only states
that can be arbitrarily shared are the separable ones. Recall that
a bipartite quantum state $\rho_{AC}$ in $\compl^d\otimes\compl^d$
is said to be $N$-shareable when it is possible to find a quantum
state $\rho_{AC_1\ldots C_N}$ in
$\compl^d\otimes(\compl^d)^{\otimes N}$ such that
$\rho_{AC_k}=\tr_{\bar k}\rho_{AC_1\ldots C_N}=\rho_{AC},\,\forall
k$. The state $\rho_{AC_1\ldots C_N}$ is then said to be an
$N$-extension of $\rho_{AC}$. The initial correlations between
subsystems $A$ and $C$ are now shared between $A$ and each of the
$N$ subsystems $C_i$, see Fig. \ref{entshar}. It is
straightforward to see that
\begin{equation}
    \rho_{AC_1\ldots C_N}=\sum_i q_i\proj{\alpha_i}\otimes
    \proj{\gamma_i}^{\otimes N}
\end{equation}
gives a valid $N$-extension of a separable state
$\rho_{AC}^s=\sum_i q_i\proj{\alpha_i}\otimes\proj{\gamma_i}$ for
all $N$. As proven by Werner, if the state is entangled, there
exists a finite $N$ where no valid extension can be found.

The second ingredient needed in what follows are the properties of
EBC. A channel $\Upsilon$ is said to be entanglement breaking when
it cannot be used to distribute entanglement. In Ref. \cite{HSR}
it was proven that the following three statements are equivalent:
(1) $\Upsilon$ is entanglement breaking, (2) $\Upsilon$ can be
written in the form $\Upsilon(\rho)=\sum_j\tr(M_j\rho)\rho_j$,
where $\rho_j$ are quantum states and $\{M_j\}$ defines a
measurement and (3) $(\one\otimes\Upsilon)\ket{\Phi^+}$ is a
separable state, where $\ket{\Phi^+}=\sum_i \ket{ii}/\sqrt d$ is a
maximally entangled state in $\compl^d\otimes\compl^d$. The
equivalence of (1) and (2) simply means that any EBC can be
understood as the measurement of the input state, $\rho$, followed
by the preparation of a new state $\rho_j$ depending on the
obtained outcome. The equivalence of (1) and (3) reflects that the
intuitive strategy for entanglement distribution where half of a
maximally entangled state is sent through the channel is enough to
detect if $\Upsilon$ is entanglement breaking.

After collecting all these results, we are now ready to prove the
following

{\bf Theorem:} Asymptotic cloning corresponds to state estimation.
Thus, $F_M=F_C$ for any ensemble of states.

{\sl Proof:} The idea of the proof is to characterize the quantum
maps $\L$ associated to asymptotic cloning machines. First of all,
note that, for any number of clones, we can restrict our
considerations to symmetric cloning machines, $\L^s$, where the
clones are all in the same state. Indeed, given a machine where
this is not the case, one can construct a symmetric machine
achieving the same fidelity $F_C(N)$, just by making a convex
combination of all the permutations of the $N$ clones \cite{note}.
Now, denote by $\L^c$ the effective cloning map consisting of,
first, the application of a symmetric machine $\L^s$ and then
tracing all but one of the clones, say the first one. The cloning
problem can be rephrased as, see Eq. (\ref{avfidcl}),
\begin{equation}
\label{asclon}
    \max_{\L^c}\sum_ip_i\bra{\psi_i}\L^c(\psi_i)\ket{\psi_i} .
\end{equation}
Note that this maximization runs over all channels that can be
written as $\L^c=\tr_{\bar 1}\L^s$. For instance, the identity
map, where $\psi\to\psi,\,\forall\psi$, does not satisfy this
constraint.

If the $N$-cloning map is applied to half of a maximally entangled
state, the resulting state,
\begin{equation}
    \rho_{AC_1\ldots C_N}=(\one_A\otimes\L^s_B)\ket{\Phi^+}_{AB},
\end{equation}
is such that, for all $i$,
\begin{equation}\label{rhoac}
    \rho_{AC_i}=(\one\otimes\L^c)\ket{\Phi^+}=\rho_{AC}.
\end{equation}
That is, the output of the $N$-cloning machine acting on half of a
maximally entangled state is a valid $N$-extension of $\rho_{AC}$.
When taking the limit of an infinite number of clones, and because
of the monogamy of entanglement, this implies that $\rho_{AC}$ has
to be separable and, thus, $\L^c$ is entanglement breaking
(\ref{rhoac}). Since any EBC can be seen as measurement followed
by state preparation, asymptotic cloning (\ref{asclon}) can be
written as \cite{note2}
\begin{equation}
    \max_{\{M_j,\phi_j\}}\sum_{i,j}
    p_i\, \tr(M_j\proj{\psi_i})\,|\braket{\psi_i}{\phi_j}|^2 ,
\end{equation}
which defines the optimal state estimation problem. Therefore,
$F_M=F_C$ for any ensemble of states. $\Box$

The same argument applies to the case in which $L$ copies of the
initial state $\ket{\psi}$ are given. The measurement  and cloning
fidelities now read, see Eqs. (\ref{avfidmeas}) and
(\ref{avfidcl}),
\begin{eqnarray}
  \bar F_M(L) &=& \sum_{i,j}
    p_i\, \tr(M_j\proj{\psi_i}^{\otimes L})\,|
    \braket{\psi_i}{\phi_j}|^2  \nonumber\\
  \bar F_C(N,L) &=& \sum_{i,k} p_i\, \frac{1}{N}\bra{\psi_i}
  \tr_{\bar k}\L(\psi_i^{\otimes L})\ket{\psi_i} .
\end{eqnarray}
Using the same ideas as in the previous Theorem, it is
straightforward to prove that
\begin{equation}\label{eqfidL}
    F_M(L)=F_C(N\to\infty,L),
\end{equation}
where $F_M(L)$ and $F_C(N,L)$ denote the optimal values of $\bar
F_M(L)$ and $\bar F_C(N,L)$, as above.

One can also extend this result to asymmetric scenarios. An
asymmetric cloning machine \cite{asclon}, given an initial input
state $\ket{\psi}$, produces $N_A$ clones of fidelity $F_C(N_A)$
and $N_B$ clones of fidelity $F_C(N_B)$. The machine is optimal
when it gives the largest $F_C(N_B)$ for fixed $F_C(N_A)$. In the
case of measurement, we are thinking of measurement strategies
where the goal is to obtain information on an unknown state
introducing the minimal disturbance. As above, we consider that a
guess for the input state is done depending on the measurement
outcome. The information vs disturbance trade-off can again be
expressed in terms of fidelities \cite{banaszek}: the information
gain is given by the overlap, $G$, between the initial and the
guessed state, while the disturbance is quantified by the overlap,
$F$, between the state after the measurement and the initial
state. A measurement is optimal when for fixed disturbance, $F$,
it provides the largest value of $G$. The optimal trade-off
between $F$ and $G$ has been derived in \cite{banaszek} for the
case in which the input ensemble consists of any pure state in
$\compl^d$ with uniform probability.

As it happens for the symmetric case, a connection between this
state estimation problem and asymmetric cloning machines can be
expected when $N_A=1$ and $N_B\to\infty$. Indeed, the previous
measurement strategy gives a possible realization of this
asymptotic and asymmetric cloning machine, not necessarily
optimal. Actually, when the input state is any pure state, with
uniform probability, the optimal measurement strategy of
\cite{banaszek} turns out to saturate the optimal cloning $1\to
N_A+N_B$ fidelities of \cite{IACFFG}, with $N_A=1$ and
$N_B\to\infty$. Now, the equality between the measurement and
asymptotic cloning fidelities in the asymmetric scenario for any
ensemble of input states can be proven using the same arguments as
above: one has to symmetrize the $N_B$ clones and then use the
connection with entanglement shareability and EBC when
$N_B\to\infty$.

From a more speculative point of view, there exist several works
relating the impossibility of perfect cloning to the no-signaling
principle, namely the impossibility of having faster-than-light
communication (see for instance \cite{gisin}). Actually, a
no-cloning theorem can be derived just from the no-signaling
principle, without invoking any additional quantum feature
\cite{ncsign}. In view of the strong connection between cloning
and state estimation, it would be interesting to study whether a
similar link between the no-signaling principle and the
impossibility of perfect state estimation could also be
established.

To conclude, this work proves the long-standing conjecture on the
equivalence between asymptotic cloning and state estimation. It
represents the strongest link between two fundamental no-go
theorems of Quantum Mechanics, namely the impossibilities of
perfect cloning and state estimation.

\medskip


We thank Emili Bagan, John Calsamiglia, Sofyan Iblisdir and Ramon
Mu\~noz-Tapia for discussion. This work is supported by the
Spanish MCyT, under ``Ram\'on y Cajal" grant, and the Generalitat
de Catalunya, 2006FIR-000082 grant.


\begin{references}
\bibitem{note}
Notice that this does not mean that the output of the cloning
machine lives in the symmetric subspace.
\bibitem{note2}
We can already restrict the guessed states to be pure, without any
loss of optimality.
\end{references}
\end{document}